\documentclass[prl,nofootinbib,%twocolumn
 ]{revtex4}

\usepackage{graphicx}
\usepackage{amssymb}
\usepackage{amsmath}
\usepackage{amsfonts}
\usepackage{epstopdf}
\usepackage{epsfig}
\usepackage{wrapfig}

\begin{document}

\title{Information is not a thermodynamic resource$^1$}{\footnote{Dedicated to the memory of Marian Smoluchowski, the most successful Maxwell demon exorcist in the 100 anniversary of his death.}
\author{Robert Alicki,}
\affiliation{Institute of Theoretical Physics and Astrophysics, University of
Gda\'nsk, Poland }

\begin{abstract}
The so-called information-thermodynamics link has been created in a series of works starting from Maxwell demon and established by the idea of transforming information into work in the though experiment of Szilard which then evolved into the vast field of research. The aim of this note is firstly to present two new models of the Szilard engine containing arbitrary number of molecules which show irrelevance of acquiring information for work extraction. Secondly, the difference between  the definition of entropy for ergodic systems and systems with ergodicity breaking constraints  is emphasized. The role of nonergodic systems as information carriers and the thermodynamic cost of stability and accuracy of information encoding and processing is briefly discussed.

\end{abstract} 
\maketitle

It is generally believed that an acquired bit of information can be traded off for a $k_B T \ln 2$ of work extracted from the bath at the temperature $T$. The most popular argument supporting this claim is a though experiment proposed by Szilard \cite{Szilard} with a single-molecule engine coupled to a single heat bath. 

In order to clarify the idea of Maxwell demon \cite{Maxwell}, so clearly and correctly exorcised by Smoluchowski \cite{Smoluchowski}, Szilard proposed a model of an engine which consists of a box, containing only a single gas particle, in thermal contact with a heat bath, and a partition.
The partition can be inserted into the box, dividing it into two equal volumes, and can slide without friction along the box. 
To extract  $k_B T\ln 2$ of work in an isothermal process of gas expansion one connects up the partition to a pulley. Szilard assumes that in order  to realize work extraction it is necessary to know  ``which side the molecule is on'' what corresponds to one bit of information.
\par
In this picture energy used to insert the partition is negligible  while the subjective lack of knowledge is treated as a real thermodynamical entropy reduced by the measurement. To avoid the conflict with the Second Law of Thermodynamics it is assumed that the reduction of entropy is compensated by its increase in the environment due to dissipation of at least $k_BT\ln 2$ work invested in the feed-back protocol of work extraction. This idea initiated a never ending discussion  about the place where the external work must be invested: in measurement \cite{Brillouin:1956}, in resetting memory \cite{Landauer:1961,Bennett:2003}, or both \cite{Sagawa:2012}.
\par
Although, the arguments in favor of the information-thermodynamics link  seem to be generally accepted, one can find in the literature several examples of doubts and criticism.\\
\begin{enumerate}
\item{
As already noticed  by Popper and Feyerabend \cite{Feyerabend}  and more recently by the author \cite{Alicki:2013}-\cite{Alicki:2014b},
one can design procedures of extracting work without knowing the position of the particle.  Various designs has been proposed to eliminate the presence of observer in work extraction.}
\item{ 
Jauch and Baron \cite{Jauch} were concerned with \emph{subjectivity} of the argument based on measurement process and stressed that the \emph{potential to do work} is present even before measurement is performed.}
\item{
A well-motivated criticism of the information-thermodynamics link has been presented by Norton \cite{Norton}. He argued (following the reasoning of Smoluchowski) that the Szilard engine cannot yield work  without the presence of external nonequilibrium reservoir. A moving piston, massive enough to suppress thermal fluctuations, can serve as such external source of work.}
\item{
Bender et.al \cite{Bender} analyzed quantum version of the Szilard engine and showed that in the energy balance one has to include work required to insert the partition (energy barrier).}
\end{enumerate}
\par
In this note two new versions of Szilard engine containing $N$ molecules are discussed, first involving measurement and feedback and  second without measurement. The generalization to $N$ molecules allows to rebut the argument that a single-molecule engine is an extreme idealization for which the laws of thermodynamics and the theory of ideal gas are not applicable.

\par
The first model is a direct extension of the Szilard design,  with $N$ molecules in the box. The length of the box is equal to $L$ and its cross-section  is put equal to one. Then, the equation of state for a  gas at the temperature $T$ reads 
\begin{equation}
p_0 L = N k_B T,
\label{gas}
\end{equation}
where $p_0$ denotes gas pressure.
\par
One assumes a frictionless motion of all mechanical elements  and  slow isothermal processes of gas compression and expansion. The work extraction cycle consists of the following steps (see Fig.1A) :

A1) Partition is inserted in the middle of the box leaving $N_R$ ($N_L = N-N_R$) molecules in the right (left) part of the box.

A2) A measurement is performed to decide in which part of the box one can find  majority of molecules, what constitutes a single bit of information.

A3) This bit of information is used to predict the direction of net force acting on the partition and extract work in isothermal process of gas expansion by connecting up the partition to a pulley.\\
\par
To compute the extracted work assume that $N_R > N_L$. Then the work is performed by the net force acting to the left
\begin{equation}
f(x) = N k_B T \Bigr(\frac{n_R}{L/2 + x} -\frac{n_L}{L/2 - x}\Bigr)
\label{force}
\end{equation}
and is given by
\begin{equation}
W = \int_0^{\ell} f(x) dx = N k_B T \Bigr[n_R \ln \bigl(1 + \frac{2\ell}{L}\bigr) + n_L \ln \bigl(1 - \frac{2\ell}{L}\bigr)  \Bigr]
\label{workA}
\end{equation}
where $x$ is the shift of partition position from the middle to the left, $\ell = L (n_R - n_L)/2$ is a maximal shift corresponding to vanishing force, $f(\ell) = 0$, and $n_{R,L}= N_{R,L}/N$. 
\par 
Inserting the expression for $\ell$ into eq.\eqref{workA} one obtains the final formula valid in the general case
\begin{equation}
W = N k_B T \Bigr[ \ln 2  - I(n_R , n_L)  \Bigr],
\label{workA1}
\end{equation}
where
\begin{equation}
I(n_R , n_L) = -(n_R \ln n_R + n_L \ln n_L)
\label{entropy}
\end{equation}
is the Shannon entropy corresponding to the probability distribution $\{n_R , n_L\}$.
\par
For $N >> 1$ the average value of $|n_R - n_L|$ is estimated as  $\sim N^{-1/2}$ and hence the average work
is given by
\begin{equation}
\bar{W} \simeq  k_B T \mathcal{O}(1).
\label{workAav}
\end{equation}
\par
The next model is again the Szilard engine with $N$ gas molecules which possesses two mechanically synchronized and symmetric pistons equipped with elastic connecting rods.  In the absence of molecules and partition the pistons could simultaneously reach position in the middle of the box.

\begin{figure}[tb]
    \centering
    \includegraphics[width=0.50\textwidth,angle=0]{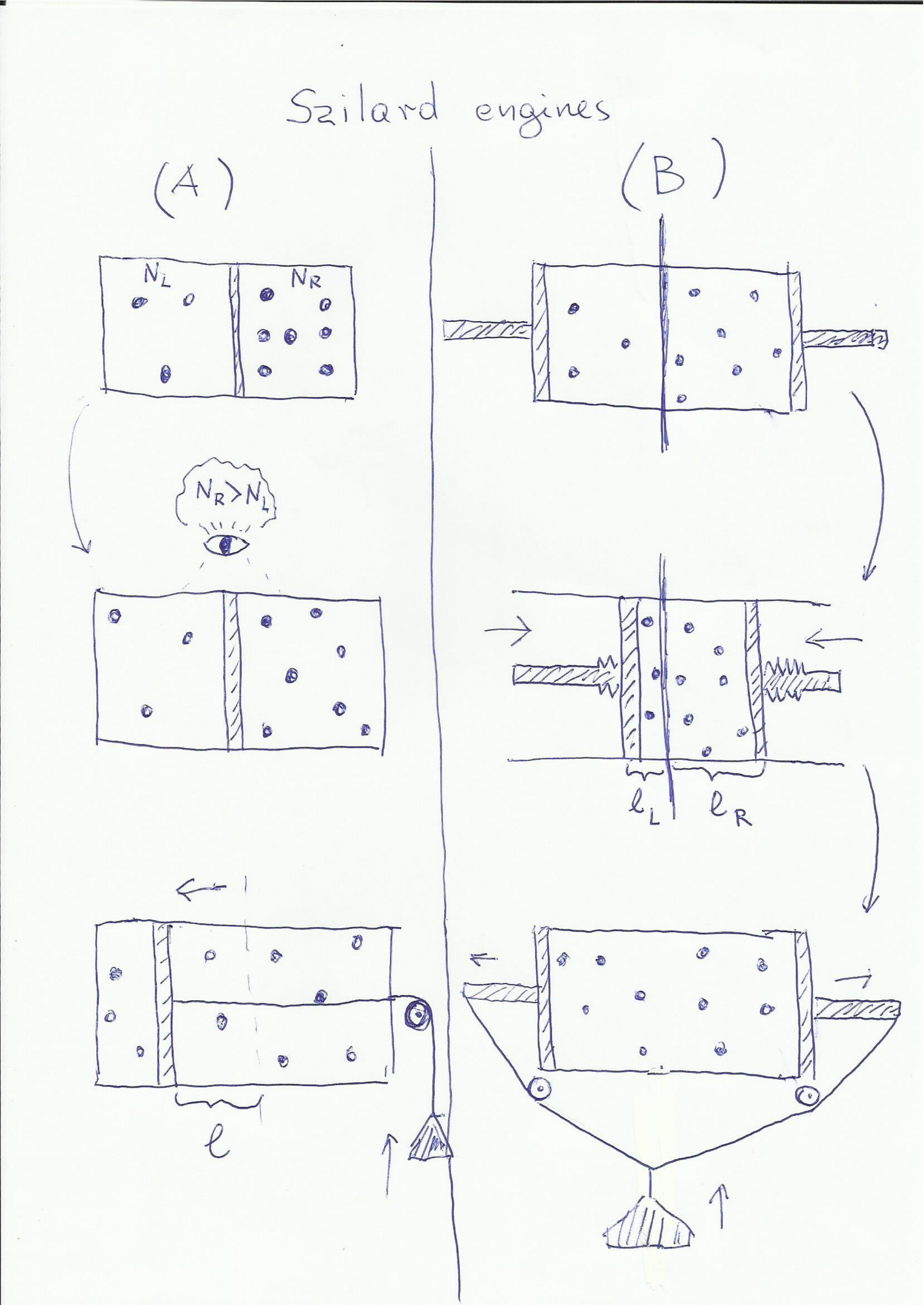}
    \caption{ Two designs of the Szilard's engine with $N$ molecules: A) involving observer, B) without observer.}
    \label{nazwa1}
 \end{figure}

\par
The work extraction cycle consists now of the following steps (see Fig.1B) :

B1) Partition is inserted. Assume that $N_R$ ($N_L$) molecule occupy the right (left) half of the box. After inserting partition the pressure acting on the right (left) piston changes from the initial value $p_0$ to $2 p_0 N_{R,L}/N $. The rods shrink or expand changing they length by $\delta L = - K^{-1} \Delta p$ where $K$ is a ``spring constant" of the rod. The corresponding potential energy change is equal to $\frac{1}{2} K (\delta L)^2 $ and scales like $K^{-1}$. 
Therefore, in the regime of large spring constant these changes of lengths and potential energies of rods can be made arbitrarily small.

B2) In the next step both pistons simultaneously move towards the middle of the box. The work needed to move each  piston is a sum of two terms. The first one $W_{R,L} (gas) = k_B T N_{R,L}\ln (L/2\ell_{R,L})$ corresponds to the isothermal compression of gas of $N_{R,L}$ molecules occupying volume $ L/2$ to  volume $\ell_{R,L}$. The second component $W_{R,L}(rod) = \frac{1}{2} K \ell_{R,L}^2$ is work needed to shrink the length of each rod by $\ell_{R,L}$. Notice, that $K \ell_{R,L} =  p_0 N_{R,L}(L/\ell_{R,L})$ while, due to \eqref{gas},  the potential energy stored in each rod is independent on $K$ and is equal to $W_{R,L}(rod) = \frac{N_{R,L}}{2} k_B T$.

B3) In the last step of the cycle the partition is removed and both pistons return to the initial external positions. The extracted work is a sum of work provided by isothermally expanding gas of $N$ particles, from the volume $\ell =\ell_R + \ell_L$ to $L$, equal to $W(gas) = k_B T\ln (L/\ell)$ and work extracted from the potential energy 
of the rods $W(rods) = W_R(rod)+ W_L(rod)$.\\  Here, again, one can neglect the change of potential energy of the rods due to gas equilibration after removal of the partition.
\par
The total work gain during the cycle is equal to
\begin{equation}
W =  \bigl[N k_B T\ln (L/\ell) + W(rods)\bigr] - \bigl[N k_B T\{n_R\ln (L/2\ell_R) + n_L\ln (L/2\ell_L)\}+ W(rods)\bigr] ,
\label{workB}
\end{equation}
\begin{equation}
W =  N k_B T\bigl[n_R\ln (2\ell_R/\ell) + n_L\ln (2\ell_L/\ell)\bigr] = N k_B T\bigl[\ln 2 - \ln(\sqrt{n_R} +\sqrt{n_L}) -\frac{1}{2} I(n_R , n_L)\bigr] .
\label{work1}
\end{equation}
Similarly to the previous model, for $N >> 1$ the averaged extracted work does not depend essentially on $N$ and can be estimated by 

\begin{equation}
\bar{W} \simeq k_B T\mathcal{O}(1).
\label{workBav}
\end{equation}
The final result is also independent on $K$, hence the assumption of large spring constant allowing to neglect the deformations of rods during inserting and removing of the partition is consistent with the estimation \eqref{workBav}. 
\par
Obviously, the second protocol of work extraction from a single heat bath does not need any presence of observer acquiring information.
Moreover, no information about the distribution of gas molecules,  when partition is inserted, remains recorded in any way at the end of the cycle. 
Therefore, the validity of the Second Law in the form : ``one cannot extract work from a single heat bath in a cyclic process" implies that
the missing work must be provided by some external source. The small value of extracted work and its asymptotic independence on $N$ does not dismiss the argument. One can combine a large number of such engines (each of the moderate size albeit large enough to apply ideal gas law) linked together to obtain a macroscopic power output \cite{Zurek}.
\par 
Because ``the potential to do work" appears just after inserting partition in the form of a non-zero net pressure, the external source of work is provided by the mechanism of partition. The partition can be treated as a ``switch" operating between two positions: ``outside" and ``inside" of the box. In order to stabilize a switch in the presence of thermal noise one needs to introduce potential barriers and friction mechanism. During its evolution between two meta-stable positions partition interacting with the gas molecule behaves like a nonequilibrium bath which can transfer free energy to gas particles. The energy scale of these processes must be much larger than the thermal energy $k_B T $, otherwise the partition would fluctuate itself suppressing the mechanism of work extraction\footnote{Compare with the analysis of Maxwell demon by Smoluchowski in terms of fluctuating trapdoor\cite{Smoluchowski}.}.
\par
One can now analyze the discussed Szilard engine from a general thermodynamic perspective. Consider a more general classical system at the thermal equilibrium  occupying a certain phase-space $X$ which can be virtually decomposed into disjoint subsets, $X = \cup_{j=1}^n X_j$. If the system is ergodic, i.e. during its evolution visits the whole phase-space its equilibrium entropy can be identified with  $ S= k_B \ln |X|$ where $|X|$ denotes the volume of $X$ \footnote{The similar analysis can be done using Gibbs canonical distributions with Boltzmann entropy. The results are the same.}. 
\par
Assume now that the decomposition is real (physical), i.e. caused by introducing potential barriers which breaks ergodicity of the system by suppressing transitions between subsets $X_j$. This action creates essentially a new physical system consisting of $n$ ergodic components, in a new equilibrium state \cite{Ishioka}. Such system can be easily distinguished from the initial one. For example, values of pressure executed on various parts of  boundary are different as well as the observed characteristics of system's time-evolution like frequencies of periodic trajectories. If the process is isothermal and change in the total energy is negligible and we can discuss only system's entropy which after physical partitioning of the phase space is given by the average over all ergodic components
\begin{equation}
S' = \sum_j \lambda_j S_j = S + k_B \sum_j \lambda_j \ln\lambda_j, \quad \lambda_j = \frac{|X_j|}{|X|}, \quad S_j = k_B \ln|X_j|.
\label{new_ent}
\end{equation}
This lower entropy $S'$ differs from the initial one $S$ by the factor $k_B I(\vec{\lambda})$ where $I(\vec{\lambda})= -\sum_j \lambda_j \ln\lambda_j$
is interpreted as information gained by the observer who detected  which part of the phase-space is actually occupied after partitioning. Moreover, it is claimed that the reduction of thermodynamical entropy from $S$ to $S'$ is a consequence of this information acquiring process.
\par
The discussion of above illustrated by the model of Szilard engine shows that the entropy reduction from $S$ to $S'$ is due to the physical  process of partitioning which costs an amount of work much larger than $k_BTI(\vec{\lambda})$. Neither observer nor measurement device are needed in this process. The consequence of entropy reduction is an increase of free energy
which can be utilized in terms of work extraction. By a proper physical procedure including temporal removal of constraints introduced for partitioning one can extract maximum $k_BT I(\vec{\lambda})$ of work in a cyclic process.
\par
The similar analysis with similar results should be possible for quantum systems. Instead of the phase-space decomposed into subsets the Hilbert space of the system is decomposed into orthogonal subspaces, $\mathcal{H} = \oplus_{j=1}^n \mathcal{H}_j $. Indeed, various quantum versions of Maxwell demon and single-molecule Szilard engine have been discussed (see e.g.\cite{Bender},\cite{Zurek}). The main conceptual problem is the typical assumption that entropy reduction from $S$ to $S'$ (see \eqref{new_ent}) happens after the measurement and is due to the ``collapse of wave function" \footnote{Bender et.al \cite{Bender} showed that such a collapse can lead to the rather unphysical instantaneous energy transfer between the ergodic components of the system.}. Reconciliation of this assumption with the Second Law leads to the hypothesis of information-thermodynamics link.
\par
On the other hand the quantum-classical correspondence and the desired lack of subjectivity suggest  that the classical concept of energy provided by the partition mechanism is correct. Further studies of quantum models with dynamically induced superselection rules and properly defined work and heat are still necessary to fully clarify this issue.
\par
Although the picture of information being a thermodynamic resource is an illusion, there exist interesting physical questions of  the fundamental nature related to information processing. Any physical system which, by means of external controlled constraints, can be decomposed into $n$ ergodic components can serve as an universal model of information carrier with the capacity of $\log_2 n $ bits. However, the constrains are never perfect because they are always realized by finite energy barriers. In the presence of thermal noise any finite barrier is penetrable and its shape determines the life-time of encoded information at the given temperature \footnote{The life-time can be estimated using Kramers theory of reaction kinetics.}. For the same reasons the system localization in a given ergodic component is also never perfect and  probability tails penetrating the other components yield the error of information encoding.
\par
The interesting problem is  the thermodynamic cost of information processing. As an elementary ingredient of information processing hardware one can take a \emph{switch} - the device which can move the  system from one ergodic component to another and back. Two models of a switch have been studied independently: a classical electronic switch by Kish \cite{Kish} and a quantum switch, described by the spin-boson model, by the author \cite{Alicki:2013}-\cite{Alicki:2014b}. In both cases minimal work necessary to change the position of a switch is approximated by the compact expression
\begin{equation}
W \simeq k_B T \ln\frac{1}{\epsilon}, 
\label{work2}
\end{equation}
where $\epsilon$ is the error of information encoding. Moreover, the life-time of encoded information is estimated by $\tau \simeq \frac{\tau_0}{\epsilon}$, where $\tau_0$ is the relaxation time of the system in the absence of ergodicity breaking constrains. The formula \eqref{work2} is valid for $\epsilon << 1$ and in the high-temperature regime for the quantum model.
It it interesting that the formula \eqref{work2} has been obtained by Brillouin a long time ago \cite{Brillouin:1956}, for a particular example of measurement involving photons. In that case $W$ was a minimal work necessary to perform this measurement with the accuracy $\epsilon$.
\par
The author thanks Philipp Strasberg for pointing out an error in the previous version of this paper, Bernhard Meister for providing useful references and Micha\l\ Horodecki for discussions.

\end{document}